\documentclass[a4paper,12pt,onecolumn,notitlepage]{article}
\usepackage[pdfpagemode=None,bookmarksopen=false,colorlinks=true,linkcolor=blue,citecolor=blue]{hyperref}
\usepackage{amsmath}
\usepackage{amssymb}
\usepackage{graphicx}
\usepackage{wrapfig}
\usepackage{sectsty}
\usepackage{bibspacing}
\usepackage{rpkid}
\allsectionsfont{\normalsize}
\begin{document}

\begin{center}
\noindent {\LARGE \textbf{Quantum Schwarzschild space-time}}\\\ \\
Pawe{\l} Duch\\
{\small\textit{Faculty of Physics, University of Warsaw, Ho\.za 69, 00-681 Warszawa, Poland}}\\{\small\texttt{pduch@fuw.edu.pl}}\\\ \\
Ryszard Pawe{\l} Kostecki\\
{\small\textit{Institute of Theoretical Physics, University of Warsaw, Ho\.za 69, 00-681 Warszawa, Poland}}\\{\small\texttt{kostecki@fuw.edu.pl}}\\\ \\
{January 11, 2012}
\end{center}

\begin{abstract}
Using new approach to construction of space-times emerging from quantum information theory, we identify the space of quantum states that generates the Schwarzschild space-time. No quantisation procedure is used. The emergent space-time is obtained by the Poincar\'{e}--Wick rotation and Fronsdal embedding of certain submanifold of the riemanian manifold of six-dimensional strictly positive matrices with the  Bogolyubov--Kubo--Mori metric.
\end{abstract}

\section{Introduction}

One of us (RPK) has recently proposed a new approach to the problem of unification of quantum theory with general relativity theory. Its key idea is to ``general relativise quantum theory'' instead of ``quantising general relativity''. The main motivation are the important conceptual and mathematical problems of main approaches to ``quantisation of gravity'', as well as the belief that quantum theory (and its unification with general relativity) requires solid conceptual and mathematical foundations free of the concept of quantisation and free of perturbative expansions.

The main tool allowing to develop this idea is the new approach to foundations of quantum theory, proposed in \cite{Kostecki:2010:AIP,Kostecki:2011:Vaxjo}. According to it, the kinematics of quantum theory is a direct extension of probability theory to the regime where measures on commutative boolean algebras are replaced by integrals on non-commutative algebras. The novel and key mathematical aspect is provided by the use of the Falcone--Takesaki non-commutative integration theory \cite{Falcone:Takesaki:2001}, which allows to construct new mathematical framework for quantum theory without relying on Hilbert spaces or measure spaces in foundations. The novel and key conceptual aspect is provided by replacement of the orthodox linear geometry of Hilbert spaces by the non-linear quantum information geometry of spaces of integrals on non-commutative $W^*$-algebras. The striking feature of this geometry is that it reduces in special cases to projective (norm) geometry of complex Hilbert spaces and to riemannian geometry of smooth differential manifolds. More generally, the new kinematics of quantum theory consists of two levels. The `non-linear' level consists of quantum models $\M(\N)$, defined as subsets of positive part of the Banach predual $\N_*$ of a non-commutative $W^*$-algebra $\N$, equipped with a non-linear quantum information geometry (which is determined by some choice of such geometric entities on $\M(\N)$ as quantum relative entropy, riemannian metric, affine connection, etc.). The `linear' level consists of representations of this geometry in terms of linear non-commutative $L_p(\N)$ spaces. In particular, the $L_2(\N)$ space can be naturally equipped with an inner product, which makes it isometrically isomorphic to the Hilbert space $\H$ (of Haagerup's standard representation). This allows for a recovery of kinematics of the orthodox approach to quantum theory as a special (self-dual) linear representation of the generically non-linear kinematics of $\M(\N)$.

This foundational framework for quantum theory offers new answers to the question ``how to reconcile quantum theory with general relativity?'', leading to a new approach to the problem of ``quantum gravity''. The quantum model $\M(\N)$ together with its quantum information geometry is considered as the main underlying kinematic object of the theory, while the space-time geometry is considered as an emergent entity that encodes some part of the quantum information geometry of $\M(\N)$. The particular form of quantum information geometry of $\M(\N)$ depends on  the definition of the experimental situation that is subjected to description and prediction in terms of this quantum theoretic model. Because our subject of consideration is ``quantum gravity'' understood as a ``general relativised quantum theory'', we will constrain the discussion of geometric structures on $\M(\N)$ to those that allow to determine a particular quantum riemannian manifold $(\M(\N),\gbold)$. The particularly important examples include: 1) the riemannian geometry canonically derived from the Norden--Sen geometry (riemannian metric $\gbold$ and a pair of affine connections that are mutually conjugate with respect to $\gbold$) that is derived from differentiation of the single quantum relative entropy functional on $\M(\N)$; 2) the solution of some variational equation determining the riemannian metric on $\M(\N)$; 3) the riemannian geometry with such riemannian metric that is invariant with respect to the action of a given group $G$ on $\M(\N)$. (Note that the last example can be considered as an extension of the theory of $W^*$-dynamical systems, allowing for more detailed analysis and specification of the spaces of states of those systems.)

While using any of the above methods, one can observe an interesting feature of quantum information geometry: the points of quantum information geometric manifold $\M(\N)$ have internal structure, and the behaviour of smooth differential objects on $\M(\N)$ depends on this structure. For example, the particular functional form of the differential geometric objects on $\M(\N)$ depends on the choice of a functional representation of $\M(\N)$ in some linear space. As a result, the additional degree of freedom is introduced into differential geometric discussions: not only the freedom of choice of representation in terms of non-linear coordinate systems on $\M(\N)$, but also the freedom of choice of functional representation in terms of operators on some linear (typically Hilbert) space. For example, if $\M(\N)$ is represented as a space of non-normalised strictly positive matrices over $\CC$, then one can vary the dimension of the representation, while keeping the same $\dim\M(\N)$. Hence, if one wants to identify the class of quantum riemannian manifolds $(\M(\N),\gbold)$ such that certain geometric quantity $A(\gbold)$ constructed from $\gbold$ satisfies some condition $C(A(\gbold))$, then this problem might be solved either by varying internal structure of $\M(\N)$ for a fixed functional form of $\gbold$, or by varying the form of $\gbold$ for a fixed $\M(\N)$, or by varying both these objects. The particular constraints of variation of these objects have to be determined by precise specification of the corresponding experimental situation.

%If the particular quantum riemannian manifold $(\M(\N),\gbold)$ is selected, then the \textit{quantum space-time} is obtained from it by the Poincar\'{e}--Wick rotation of riemannian metric $\gbold$ to the pseudo-riemannian metric $\tilde{\gbold}$. This rotation requires to specify a field $\mathbf{e}$ of differential one-forms that provides a global foliation of $(\M(\N),\gbold)$. Thus, \textit{quantum space-times} $(\M(\N),\tilde{\gbold})$ can be obtained only from such quantum riemannian geometries $(\M(\N),\gbold)$ that allow global foliations. By the means of this construction, all quantum space-times are time-oriented globally hyperbolic pseudo-riemannian manifolds. The recovery of an ordinary general-relativistic space-times from quantum space-time amounts to forgetting about the internal structure of the quantum model $\M(\N)$. This is formalised in terms of the forgetful functor $\mathrm{DeQuant}$ from the category $\mathbf{psrQMod}$ of quantum pseudo-riemannian models $(\M(\N),\tilde{\gbold})$ with isometric embeddings as arrows to the category $\mathbf{psrMan}$ of smooth pseudo-riemannian manifolds with isometric embeddings as arrows.

If the particular quantum riemannian manifold $(\M(\N),\gbold)$ is selected, then the \textit{quantum space-time} can be obtained from it by the Poincar\'{e}--Wick rotation of riemannian metric $\gbold$ to the lorentzian metric $\tilde{\gbold}$. This requires to specify globally defined smooth field $\mathbf{e}$ of differential one-forms, and provide a decomposition
\begin{equation}
	\gbold=\mathbf{e}\otimes\mathbf{e}+\hat{\gbold},
\end{equation}
where $\hat{\gbold}$ is a riemannian metric orthogonal to $\mathbf{e}\otimes\mathbf{e}$. The Poincar\'{e}--Wick rotation amounts to substitution of riemannian metric $\gbold$ by the lorentzian metric
\begin{equation}
	\tilde{\gbold}:=-\mathbf{e}\otimes\mathbf{e}+\hat{\gbold}.
\end{equation}
The smooth vector field $Z$ defined by $\tilde{\gbold}(Z,\cdot)=\mathbf{e}$ is naturally timelike ($\tilde{\gbold}(Z,Z)<0$), so the lorentzian manifold $(\M(\N),\tilde{\gbold})$ is time-oriented. Thus, according to the standard definition \cite{Chrusciel:Galloway:Pollack:2010}, it is a space-time. We will call pairs $(\M(\N),\tilde{\gbold})$ \textit{quantum space-times}. The recovery of the ordinary space-times from quantum space-times amounts to forgetting about the internal structure of the points of quantum model $\M(\N)$. This can be formalised by introducing the forgetful functor $\mathrm{DeQuant}^{\mathrm{lor},\mathrm{to}}$ from the category $\QMod^{\mathrm{lor},\mathrm{to}}$ of time-oriented lorentzian quantum models $(\M(\N),\tilde{\gbold})$ with isometric embeddings as arrows to the category $\catname{SpaceTime}$ of space-times with isometric embeddings as arrows. 

Note that it is possible to apply the steps of the Poincar\'{e}--Wick rotation and dequantisation in the reverse order, without changing the result of the procedure. In such case, however, the notion of quantum space-time does not appear. Moreover, the problem of analytic continuation in time variable seems to find much better environment on the level of quantum models. Hence, if some additional structures (such as glbal hyperbolicity) are also required to emerge from information geometry of quantum models, then it seems reasonable to leave forgetful dequantisation as a last step of the `space-time emergence' procedure.

The goal of this paper is to use the above general framework to construct a family of quantum models $\M(\N)$ that generate a particular (Schwarzschild) class of space-times, for a particular (Bogolyubov--Kubo--Mori) class of a quantum riemannian metrics $\textbf{g}$ on $\M(\N)$. We begin by constructing a family of quantum models with elements belonging to the space of two-dimensional non-normalised strictly positive matrices that corresponds to three-dimensional flat euclidean space. Then we glue them to obtain the class of quantum models corresponding to six-dimensional euclidean space. Next, we chose a global smooth one-form field $\mathbf{e}$, and provide the Poincar\'{e}--Wick rotation of $\gbold$ with respect to $\mathbf{e}$, which results in the flat quantum space-time $(\M_6,\tilde{\gbold})$. Finally, we use the Fronsdal embedding \cite{Fronsdal:1959} of the Schwarzschild space-time to six-dimensional flat space in order to specify a manifold of quantum states that determines the Schwarzschild space-time $(\M_S,\tilde{\gbold}|_{\M_S})$.

\section{Riemannian BKM manifolds of quantum states}

\subsection{Finite dimensional quantum models over type I factor algebras}

If the $W^*$-algebra $\N$ contains no type III factor and if $\N_*^+$ contains at least one faithful element $\omega$ (i.e., $\omega(x^*x)=0\Rightarrow x=0\;\forall x\in\N$), then $\M(\N)$ can be represented as a space $\M(\H_\omega)$ of positive operators over Hilbert space $\H_\omega$. The space $\H_\omega$, as well as the representation $\pi_\omega:\N\ra\BBB(\H_\omega)$, are uniquely constructed from a pair $(\omega,\N)$ by means of the Gel'fand--Na\v{\i}mark--Segal (GNS) construction \cite{Gelfand:Naimark:1943,Segal:1947:irreducible}. If $\dim\M(\H_\omega)=d<\infty$, then $\M(\H_\omega)$ is just a subspace of the space $M_d(\CC)^+$ of $d$-dimensional \textit{non-normalised} density operators (positive matrices). Note that the `reference' faithful quantum state $\omega\in\N_*^+$ is not required to belong to $\M(\N)$.

The space $L_1(\N)$ is always isometrically isomorphic to the Banach predual $\N_*$ of $\N$, while the space $L_\infty(\N)$ is always isometrically isomorphic to $\N$ itself. If $\N$ is a type I factor, then $\pi_\omega(\N)\cong\BBB(\H_\omega)$, and the Falcone--Takesaki non-commutative $L_p(\N)$ spaces turn to the spaces $L_p(\BBB(\H_\omega),\Tr)$ of $p$-th Schatten-class operators, where $\Tr$ is a canonical trace on $\BBB(\H_\omega)$. In particular, the $L_2(\BBB(\H_\omega),\Tr)$ space is just a Hilbert space $\H_{HS}$ with the Hilbert--Schmidt scalar product $\langle A,B\rangle_{HS}:=\Tr(B^*A)$ and vectors $A,B$ provided by the elements $x$ of $\BBB(\H_\omega)$ satisfying $(\Tr(x^*x))^{1/2}<\infty$.

The embeddings of quantum models $\M(\N)$ into non-commutative $L_p(\N)$ space is provided, for $p\in[1,\infty[$, in terms of the embeddings
\begin{equation}
	\ell_{1/p}:\M(\N)\ni\phi\mapsto p\phi^{1/p}\in L_p(\N).
\label{gamma.embeddings}
\end{equation}
When $\N$ is a type I factor, this turns to the family of embeddings
\begin{equation}
	\M(\H_\omega)\ni\rho\mapsto p\rho^{1/p}\in L_p(\BBB(\H_\omega),\Tr).
\end{equation}
For $p=2$, this turns to the embedding $\rho\mapsto 2\rho^{1/2}$ of a space of non-normalised density matrices into a Hilbert space $\H_{HS}$. Extension of the above family of embeddings to the $p=\infty$ case is provided by logarithmic coordinates
\begin{equation}
	\ell_0:\M(\H_\omega)\ni\rho\mapsto\log\rho\in L_\infty(\BBB(\H_\omega),\Tr)\cong\BBB(\H_\omega).
\end{equation}

As showed by Jen\v{c}ov\'{a} \cite{Jencova:2006,Jencova:2010}, quantum models $\M(\N)$ can be equipped with differential manifold structure if all elements of $\M(\N)$ are faithful. For type I algebra $\N$ and $\dim\M(\N)=d<\infty$ this condition amounts to requiring strict positivity of elements of $\M(\H_\omega)$. This restricts considerations to the space $M_d(\CC)^+_0$ of strictly positive $d$-dimensional matrices. 

The quantum differential manifold $\M(\N)$ can be equipped with various differential geometric structures. In particular, one can consider riemannian metrics on it. Quantum information theory allows to impose some additional conditions on these metrics. The standard condition is the monotonicity of metrical distance $d_{\gbold}$ of $\gbold$ under unit-preserving ($T(\mathbb{I})=\mathbb{I}$) completely positive maps $T$,
\begin{equation}
	d_\gbold(\phi,\omega)\geq d_\gbold(\phi\circ T,\omega\circ T).%,
\label{markov.monotone.metric}
\end{equation}
Condition (\ref{markov.monotone.metric}) can be interpreted as a requirement that the loss of information content of quantum states should not lead to increase of their distinguishability. This condition selects a wide class of the Morozova--Chentsov--Petz quantum riemannian metrics \cite{Morozova:Chentsov:1989,Petz:1996:monotone,Hansen:2006}. An additional condition that $\gbold$ should allow a pair $(\nabla,\nabla^\star)$ of Norden--Sen conjugate affine connections \cite{Norden:1945,Sen:1944},
\begin{equation}
	\gbold(\nabla_uv,w)+\gbold(v,\nabla^\star_uw)=u(\gbold(v,w))\;\;\forall u,v,w\in\mathbf{T}\M(\N),
\end{equation}
selects for $\dim\M(\N)<\infty$ a family of $\gamma$-metrics, where $\gamma\in[0,1]$ for
$\M(\N)=\N_*^+$ \cite{Hasegawa:Petz:1997,Hasegawa:2003,Grasselli:2004}, and $\gamma\in\{0,1\}$
for
$\M(\N)=\N_{*1}^+:=\{\omega\in\N
_*^+\mid\omega(\mathbb{I})=1\}$ \cite{Grasselli:Streater:2001:unique:AIP,Grasselli:Streater:2001:unique:IDAQ}. For
$\gamma\in\{0,1\}$, the $\gamma$-metrics are known as the Bogolyubov--Kubo--Mori (BKM)
metrics \cite{Bogolyubov:1961,Kubo:1957,Mori:1956}, while for $\gamma\in]0,1[$ they are known as the
Wigner--Yanase--Dyson (WYD) metrics \cite{Wigner:Yanase:1963}. All $\gamma$-metrics, together with their
corresponding Norden--Sen dually flat pairs of affine connections
$(\nabla^{\gamma},\nabla^{1-\gamma})$, can be derived, for
$\dim\M(\N)<\infty$ and
$\M(\N)\subseteq\N_{*1}^+$, by
differentiation of the Hasegawa relative entropy
\begin{equation}
\begin{split}
       &D_\gamma(\omega,\phi):=\frac{1}{\gamma(1-\gamma)}\Tr(\rho_\omega-\rho_\omega^\gamma\rho_\phi^{1-\gamma})=
      \frac{1}{\gamma(1-\gamma)}-\Tr(\ell_{\gamma}(\rho_\omega)\ell_{1-\gamma}(\rho_\phi)).
\label{Hasegawa}
\end{split}
\end{equation}
This derivation is provided by \cite{IJKK:1982,Eguchi:1985,Hasegawa:1993,Lesniewski:Ruskai:1999,Jencova:2004:entropies}
\begin{equation}
	\begin{array}{rl}
		\gbold^\gamma_\phi(u,v)&:=(\partial_u)_\phi(\partial_v)_\omega D_\gamma(\phi,\omega)|_{\omega=\phi},\\
		\gbold^\gamma_\phi((\nabla^\gamma_u)_\phi v,w)&:=-(\partial_u)_\phi(\partial_v)_\phi(\partial_w)_\omega D_\gamma(\phi,\omega)|_{\omega=\phi},\\
		\gbold^\gamma_\phi(v,(\nabla_u^{1-\gamma})_\phi w)&:=-(\partial_u)_\omega(\partial_w)_\omega(\partial_v)_\phi D_\gamma(\phi,\omega)|_{\omega=\phi},
	\end{array}
\label{quantum.Eguchi}
\end{equation}
where $(\partial_u)_\phi$ is a directional derivative at $\phi$ in the direction $u\in\mathbf{T}\M(\N)$. In particular, the BKM metric follows from differentiation of the Umegaki relative entropy \cite{Umegaki:1962}
\begin{equation}
	D_1(\phi,\omega):=\Tr(\rho_\phi(\log\rho_\phi-\log\rho_\omega)),
\end{equation}
which is a $\gamma\ra 1$ limit of a Hasegawa relative entropy. Taking into account the key role played by the Umegaki relative entropy in quantum information theory (as opposed to other Hasegawa relative entropies), as well as the uniqueness of the BKM metric as the only monotone riemannian metric with flat Norden--Sen dual connections on the space of normalised quantum states \cite{Grasselli:Streater:2001:unique:IDAQ,Grasselli:Streater:2001:unique:AIP}, we will restrict our considerations to the BKM quantum riemannian metrics.

It is important to note that the vectors of tangent space $\mathbf{T}_\phi\M(\N)$ admit different representation in terms of $L_p(\N)$ spaces, corresponding to the various embeddings (\ref{gamma.embeddings}). Depending on the choice of particular representation of the tangent space of $\M(\N)$, the particular quantum riemannian metric $\gbold$ can take different functional forms. Equation (\ref{Hasegawa}) shows that the choice of a particular $\gamma$-metric leads to a natural choice of a preferred pair of coordinate systems, associated with a preferred non-commutative $L_p(\N)$ space representation via $p=1/\gamma$. For this reason, we will consider the BKM metric expressed in terms of the logarithmic coordinates.

\subsection{The logarithmic representation of the BKM metric}

The mapping
\begin{equation}
 \ell_0\equiv\log : M_d(\mathbb{C})^{+}_0\ni\rho\mapsto\log\rho\in M_d(\mathbb{C})^{\textrm{sa}}
\label{log.coord}
\end{equation}
is a diffeomorphism acting on a space $M_d(\CC)^+_0$ of $d$-dimensional strictly positive matrices to a space $M_d(\CC)^{\textrm{sa}}$ of $d$-dimensional hermitean matrices. In what follows, we will consider the submanifolds of $M_d(\CC)^{\textrm{sa}}$ that are obtained using this mapping. In particular, an $d$-dimensional submanifold $\mathcal{Q}_d$ of hermitean matrices corresponds to $d$-dimensional submanifold $\exp(\mathcal{Q}_d)$ of strictly positive matrices.

If the mapping  
\begin{equation}
 H:~\mathbb{R}^d \supset \mathcal{O} \ni x^a \mapsto H(x^a) \in \mathcal{U} \subset \mathcal{Q}_d
\label{parametryzacja}
\end{equation}
is a diffeomorphism of open subsets $\mathcal{O} \subset \mathbb{R}^d$ and $\mathcal{U} \subset \mathcal{Q}_d$, then it is called a \textit{parametrisation} of an open subset $\mathcal{U}$ of a manifold $\mathcal{Q}_d$. The inverse map $H^{-1}$ %
%\begin{equation}
% H^{-1}:~\mathcal{Q}_d \supset \mathcal{U} \ni y^a \mapsto H^{-1}(y^a) \in \mathcal{O} \subset \mathbb{R}^d,
%\label{mapa}
%\end{equation}
is called a \textit{coordinate system} on $\mathcal{U}$.

Using the $\log$ map (\ref{log.coord}), we can identify $M_d(\mathbb{C})^{+}_0$ with $M_d(\mathbb{C})^{\textrm{sa}}$. We can introduce the Bogolyubov--Kubo--Mori product directly on submanifold $\mathcal{Q}_d$ of $M_d(\CC)^{\textrm{sa}}$:
\begin{equation}
 \gbold_{h}(A,B) = \int_0^1~ \dd \alpha \Tr\left\{ \mathrm{e}^{\alpha h} A \mathrm{e}^{(1-\alpha) h} B \right\},
\label{BKM.log.param}
\end{equation}
where $h \in \mathcal{Q}_d$, while
$A,B \in \mathbf{T}_h\mathcal{Q}_d \subset \mathbf{T}_h M_d(\mathbb{C})^{\textrm{sa}}\cong M_d(\mathbb{C})^{\textrm{sa}}$.

Using the parametrisation $H$ of $\mathcal{Q}_d$ defined by (\ref{parametryzacja}), we can locally express matrix elements of the metric tensor (\ref{BKM.log.param}) as
\begin{equation}
 \gbold_{ab}(x^c) = \int_0^1~ \dd \alpha \Tr\left\{ \mathrm{e}^{\alpha H(x^c)}  (\partial_a H)(x^c) 
                                                \mathrm{e}^{(1-\alpha) H(x^c)} (\partial_b H)(x^c) \right\}.
\label{BKM.posr}
\end{equation}
This can be simplified to
\begin{equation}
 \gbold_{ab}(x^c) = \partial_a \partial_b \Tr\left\{   \mathrm{e}^{H(x^c)}  \right\} 
               -  \Tr\left\{ (\partial_a \partial_b H)(x^c)  \mathrm{e}^{H(x^c)}  \right\},
\label{wzor_g}
\end{equation}
which follows from the equations:
\begin{equation}
\begin{split}
\partial_a  \mathrm{e}^H&=\int_{0}^1 \dd\alpha  \mathrm{e}^{\alpha H} (\partial_a H)  \mathrm{e}^{(1-\alpha) H},\\
 %\partial_a \Tr\{ \e^H \} &=  \int_{0}^1 \dd \alpha ~ 
  %  \Tr\left\{  \e^{\alpha H} (\partial_a H)  \e^{(1-\alpha) H} \right\}
   %  =\Tr\left\{ \e^H (\partial_a H) \right\},  \\
  \partial_a\partial_b \Tr\left\{   \mathrm{e}^H \right\} &= 
    ~\partial_a \Tr\left\{    \mathrm{e}^{H}\partial_b H \right\} = \\
  &=\int_{0}^1 \dd\alpha \Tr
    \left\{  \mathrm{e}^{\alpha H} (\partial_a H)  \mathrm{e}^{(1-\alpha) H} (\partial_b H) \right\}
    +  \Tr\left\{   \mathrm{e}^H\partial_a \partial_b H \right\}= \\
  &= ~ \gbold_{ab} +\Tr\left\{ (\partial_a \partial_b H)  \mathrm{e}^{H}  \right\}.
\end{split}
\end{equation}

\subsection{Family $Q_{fg}$ of 3-dimensional riemannian manifolds}

Let
\begin{equation}
\begin{split}
\mathcal{F} := \big\{ (f,g) \in  C^\infty(\mathbb{R}_+,\mathbb{R})^2 
                        ~\big|~~\forall_{k\in \mathbb{N}}~ f^{(2k)}(0)=0,~ g^{(2k+1)}(0)=0,& \\
   \mathbb{R}_+ \ni r \mapsto (f(r),g(r)) \in \mathbb{R}^2 \textrm{ is injective immersion}\big\} & 
\end{split}
\end{equation}
and
\begin{equation}
 H_{f g}: \mathbb{R}^3 \ni  \vec{x} \mapsto H_{f g}(\vec{x}) 
              := \sum_{a=1}^3 f(|\vec{x}|) \frac{x^a}{|\vec{x}|} \sigma^a 
                  + g(|\vec{x}|) \in M_2(\mathbb{C})^{\textrm{sa}},
\label{H_map}
\end{equation}
where $|\vec{x}| := \sqrt{(x^1)^2+(x^2)^2+(x^3)^2}$, and $\sigma^a$ are Pauli matrices. Consider the family $Q_{fg}$ of submanifolds of the space $M_2(\mathbb{C})^{\textrm{sa}}$ of 2-dimensional hermitean matrices, defined by
\begin{equation}
 \mathcal{F} \ni (f,g) \mapsto Q_{f g} := 
               H_{f g}\left( \mathbb{R}^3 \right) \subset M_2(\mathbb{C})^{\textrm{sa}},
\end{equation}
By definition, 
$\mathcal{F}$ is the largest set of functions $(f,g)$ for which the manifold $Q_{f g}$ is a smooth submanifold of $M_2(\mathbb{C})^{\textrm{sa}}$.

Every element of the family $Q_{f g}$ is equipped with a natural global parametrisation $H_{f g}$ and is diffeomorphic to
$\mathbb{R}^3$. The manifold $Q_{f g}$ can be parametrised by spherical coordinates $(r,\theta,\phi)$, which are introduced from $\mathbb{R}^3$ by the inverse map $\tilde{H}^{-1}_{fg}$, where
\begin{equation}
\begin{split}
 \tilde{H}_{f g}:~ \mathbb{R}_+ \times [0,\pi[ \times [0,2 \pi[\;\ni (r,\theta,\phi)  
                 \mapsto \tilde{H}_{f g}(r,\theta,\phi)\in Q_{f g},\\ 
\tilde{H}_{f g}(r,\theta,\phi) = 
                          f(r) \left( \sigma^1 \sin\theta \cos\phi +
                                      \sigma^2 \sin\theta \sin\phi  +
                                      \sigma^3 \cos\theta \right)
                          + g(r).  
\end{split}
\end{equation}

The non-zero matrix elements of the BKM metric tensor $\gbold$, calculated in these coordinates using (\ref{wzor_g}) read
\begin{equation}
\left\{
 \begin{array}{l}
  \gbold_{rr} = 2 e^{g(r)} \left(2 f'(r) g'(r) \sinh f(r) +
   \left(f'(r)^2+g'(r)^2\right)\right) \cosh f(r)\\
  \gbold_{\theta \theta} = 2 f(r) e^{g(r)} \sinh f(r)\\
  \gbold_{\phi \phi} = 2 f(r) e^{g(r)} \sin^2(\theta) \sinh f(r).\\
 \end{array}
\right.
\label{Q_BKM}
\end{equation}

\section{Schwarzschild space-time of quantum states}

\subsection{The flatness condition and its particular solution}

The manifold $Q_{f g}$ is flat if

\begin{equation}
 \gbold_{rr} = 1,\;\;\gbold_{\theta \theta} = r^2,\;\; \gbold_{\phi \phi} = r^2 \sin^2(\theta).
\end{equation}
The converse statement does not hold in general. Equations (\ref{Q_BKM}) and the above conditions give
\begin{equation}
\left\{
 \begin{array}{l}
  2 e^{g(r)} \left(  2 f'(r) g'(r)\sinh(f(r))+\cosh (f(r)) \left(f'(r)^2+g'(r)^2\right)\right) = 1, \\
  2 f(r) e^{g(r)} \sinh (f(r)) = r^2.\\
\end{array}
\right.
\label{Q_BKM_flat}
\end{equation} 
In what follows we will construct the solutions $(f,g)\in\mathcal{F}$ of (\ref{Q_BKM_flat}).

The above system of differential equations is equivalent to
\begin{equation}
\left\{
 \begin{array}{l}
   g(r) = \log \left( \frac{r^2}{2 f(r) \sinh f(r)} \right) \\     
   Q'(r) f'(r) \sinh f(r) + 
       \frac{r^2 f'(r)^2 \cosh f(r)}{2 f(r) \sinh f(r)} +
    Q'(r)^2 
      \frac{f(r)\sinh f(r) \cosh f(r) }{2 r^2} = \frac{1}{2}. 
\end{array}
\right.
\label{Q_BKM_flat2}
\end{equation}
where $Q(r)=\left[ \frac{r^2}{f(r) \sinh f(r)} \right]$.
The second equation in (\ref{Q_BKM_flat2}) is quadratic with respect to $f'(r)$, so we can replace it by one of the equations of the form
\begin{equation}
 r f'(r) = F_a(f(r)),\\
 \label{Q_BKM_flat3}
\end{equation}
where $F_a$ for $a=\pm1$ are odd functions that are analytic in some neighborhood of the real line
\begin{equation}
  F_\pm(f) = f ~ \frac{ 2 \left( 1 + \frac{f \tanh f}{\sinh^2 f } \right) \pm
                     \sqrt{f \tanh f  \left(1 + \frac{f^2}{\sinh^2 f } - \frac{2 f \tanh f}{\sinh^2 f } \right)}}
               {1 + \frac{f^2}{\sinh^2 f } + \frac{2 f \tanh f}{\sinh^2 f}}.
\end{equation}
Behaviour of $F_-$ and $F_+$ is presented on Figure 1. The function $F_+(f)$ has one root in $f=0$, while $F_-(f)$ has roots in $f=0$ and in $f=\pm f_r$ ($f_r>0$). Moreover, $\lim_{f \rightarrow \infty}\frac{F_+(f)}{f^{3/2}}=1$ and $\lim_{f \rightarrow \infty}\frac{F_\pm(f)}{f}=1$.

\begin{figure}[h!]
\begin{center}
  \includegraphics[width=0.45\textwidth]{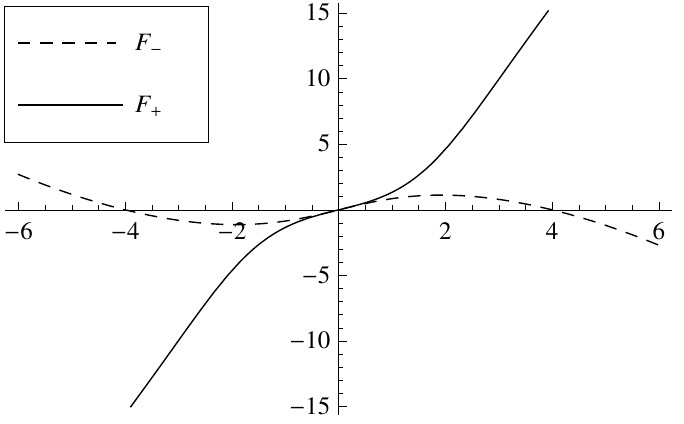}
\end{center}
\caption{The plot of functions $F_+$ and $F_-$.}
\end{figure}

According to first equation of (\ref{Q_BKM_flat2}), $g$ is unambiguously determined by $f$, so for $(f,g)\in\mathcal{F}$, $f$ has to be invertible. Hence, one can consider $r$ as a function of $f$. Then (\ref{Q_BKM_flat3}) is equivalent to
\begin{equation}
 r'(f) = r(f) \frac{1}{F_a(f)},\\
 \label{Q_BKM_flat4}
\end{equation}
The local solution of this equation for $f>0$ is given by
\begin{equation}
 r_{a,C,f_0}(f) = C \exp\left( \int_{f_0}^f \frac{\dd y}{F_a(y)} \right),
\end{equation}
where $a=\pm1$ while $f_0>0$ and $C > 0$ are constants. In case of $a=-1$, we assume also that $f_0<f_r$. The assumptions $f_0>0$ and $f_0<f_r$ follow from the choice of the interval where the function $1/F_-(f)$ does not have poles. From the fact that both $F_-(f)$ and $F_+(f)$ have a root at $f=0$ and are analytic, it follows that $1/F_-(f)$ and $1/F_+(f)$ have a pole in $f=0$. Hence, $\lim_{f\rightarrow 0} r_{a,C,f_0}(f) =0$ for $a=\pm1$. Similar reasoning leads to conclusion that 
\[
	\lim_{f\rightarrow f_r} r_{-,C,f_0}(f) =\infty.
\]
For $a=+1$ the latter conclusion does not hold. From the asymptotic behaviour of $F_2$ we obtain
\[
 \lim_{f\rightarrow \infty} r_{+,C,f_0}(f)=: r_{+,\textrm{max}} <\infty.
\]

The functions
\begin{equation}
\begin{split} 
   r_{-,C,f_0}:~ ]0,f_r[ \ni f \mapsto r_{-,C,f_0} \in \mathbb{R}_+,\\
   r_{+,C,f_0}:~ ]0,\infty[ \ni f \mapsto r_{+,C,f_0} \in ]0,r_{+,\textrm{max}}[,
\end{split} 
\end{equation}
are bijections regardless of the choice of constants $C>0$ and $f_0>0$ (in case of $a=-1$ we require $f_0\in]0,f_r[$), because $F_-(f)>0$ for $f\in ]0,f_r[$ and $F_+(f)>0$ for $f\in ]0,\infty[$ imply that $r_{a,C,f_0}$ is strictly increasing. In what follows, we will assume $a=-1$, because we want the inverse of $r_{a,C,f_0}$ to be defined globally on $\mathbb{R}_+$. We could also choose $f_0<0$, thereby obtaining a solution of (\ref{Q_BKM_flat4}) for negative $f$ only, which is the mirror reflection of $r_{-,C,-f_0}$.

It remains to show that $(f,g)\in\mathcal{F}$ for $f=r^{-1}_{-,C,f_0}$, where $C>0$, $f_0 \in ]0,f_r[$ are fixed constants, while
\[
	g(r) = \log\left( \frac{r^2}{2 f(r) \sinh(f(r))} \right).
\]
The non-trivial part of the proof amounts to showing that $\forall_{k\in \mathbb{N}}~ f^{(2k)}(0)=0,~ g^{(2k+1)}(0)=0$. It is equivalent to the smoothness of manifold in $r=0$. Below we outline the necessary steps of the proof:
\begin{enumerate}
\item From $\lim_{f \rightarrow 0} \frac{F_-(f)}{f}=1$ and (\ref{Q_BKM_flat4}) it follows that $\lim_{r \rightarrow 0}  \frac{f(r)}{r}$ exists.
\item It follows by induction that $f^{(n)}(r)$ has the form $\frac{1}{r^n}K_n(f(r))$ for $n=0,1,2,\ldots$, where $K_n(f)$ are odd analytic functions of $f$ which have zeros of $2\lfloor \frac{n}{2} \rfloor+1$ order in $f=0$. For $n=1$ we have $K_1=F_-$. In the proof of inductive step one needs to use equation (\ref{Q_BKM_flat3}) for $a=1$, the fact that only odd coefficients of power series of $K_n$ are non-zero and $\lim_{f \rightarrow 0} \frac{F_-(f)}{f}=1$. 
\item Because of step one and two all derivatives of $f$ exist at $r=0$ and fulfill desired properties. 
\item The function $g$ is well defined at $r=0$ and
\[
g'(r) = \frac{1}{r} G(f(r)),
\]
where
\[
	G(f) := 2- \left(\coth(f) + \frac{1}{f}\right)F_-(f)
\]
is an even analytical function of $f$  which has a zero of second order at $f=0$.
\item $g^{(n)}(r)$ has the form $\frac{1}{r^n} L_n(f(r))$ for $n=1,2,\ldots$, where $L_n(f)$ are even analytic functions of $f$ which have zeros of $2\lceil \frac{n}{2} \rceil$ order in $f=0$. It follows by induction similarly as in step two (note that $L_1=G$).
\item From the two preceding steps it follows that all derivatives of $g$ take finite values at $r=0$ and have required properties.
\end{enumerate}
We conclude that $(f,g)\in\mathcal{F}$, which finishes the construction of a three-dimensional flat manifold $Q_{fg}$.

\subsection{Construction of the quantum Schwarzschild space-time}

Now we are ready to construct quantum Schwarzschild space-time, as a particular four-dimensional submanifold in six-dimensional flat manifold of four-dimensional hermitean matrices (corresponding by (\ref{log.coord}) to the manifold of four-dimensional strictly positive matrices).

Let us chose any of the functions $f$ determined in the previous section (this is done by choosing $C$, $f_0\in]0,f_r[$ and setting $f=r_{-,C,f_0}^{-1}$), and define $g(r) = \log\left( \frac{r^2}{2 f(r) \sinh(f(r))} \right)$. Using the map $H_{fg}$ given by (\ref{H_map}), we define the following smooth injection
\begin{equation}
 H_6: \mathbb{R}^6 \cong \mathbb{R}^3 \oplus \mathbb{R}^3 \ni (\vec{x},\vec{y}) \mapsto 
                       H_{f g} (\vec{x}) \oplus H_{f g} (\vec{y}) \in M_4(\mathbb{C})^{\textrm{sa}}.
\label{H6_map}
\end{equation}
Let $\M_6 := H_6(\mathbb{R}^6)$. The map $H_6$ is a diffeomorphism between
$\mathbb{R}^6$ and $\M_6$. The space $(\M_6,\gbold)$, where $\gbold$ is a BKM metric on $\M_6$, is a riemannian manifold isometric (by $H_6$) to a $6$-dimensional euclidean space.

We define a one-form field by $\mathbf{e}:=\mathrm{d}x^1$. As a result of the Poincar\'{e}--Wick rotation of riemannian manifold $(\M_6,\gbold)$ with respect to $\mathbf{e}$, we obtain a flat pseudo-euclidean manifold, denoted by $(\M_6,\tilde{\gbold})$. The signature of $\tilde{\gbold}$ is $(-,+,+,+,+,+)$.

Now we can use the Fronsdal \cite{Fronsdal:1959} embedding of Schwarzschild space-time to $6$-dimensional pseudo-euclidean space
\begin{equation}
%\begin{split}
  \M_S =%&
   \bigg\{ (\vec{x},\vec{y}) \in \M_6 \bigg|%\\
    %&
    \left(x^2\right)^2 -  \left(x^1\right)^2 = 16~m^2 \left( 1- 2 m/|\vec{y}| \right),~
    x^3=h(|\vec{y}|)\bigg\}   
%\end{split} 
\label{Schwarz.big}
\end{equation}
where we implictly use parametrization $H_6$ of $\M_6$ and introduce function 
\[
	h(y) = \int_{2m}^{y} \dd r [(2m r^2 +4m^2 r + 8m^3)/r^3]^{1/2}.
\]
The constant $m>0$ is a mass parameter characterising the solution. The space $(\M_S,\tilde{\gbold}|_{\M_S})$ is a maximal extension of the Schwarzschild space-time, known as the Kruskal--Szekeres  extension \cite{Kruskal:1960,Szekeres:1960}. Instead of $\M_S$, one can also choose the manifold
\begin{equation}
\begin{split}
 \M_S' =& \bigg\{ (\vec{x},\vec{y}) \in \M_6 \bigg|\\
    &\left(x^2\right)^2 -  \left(x^1\right)^2 = 16~m^2 \left( 1- 2 m/|\vec{y}| \right),~x^1+x^2>0,~
x^3=h(|\vec{y}|)\bigg\}  
\end{split} 
\end{equation}
which corresponds to the region of Schwarzschild solution considered originally in \cite{Schwarzschild:1916}. Then the limit $m\rightarrow 0$ of $\M_S'$ is just a (flat) Minkowski space-time.

\section{Discussion}

In the preceding section we have shown that the quantum Schwarzschild space-time can be constructed as a result of particular choices of: 1) a manifold of non-normalised strictly positive density matrices, 2) a metric tensor on this space, and 3) global smooth field of one-forms (which defines the time orientation). All these choices are of purely kinematic character. When provided, they establish the emergence of a particular space-time from quantum information data.

Recall that quantum models $\M(\N)$ can be considered as manifolds if they consist of faithful elements only. In the case of models over finite-dimensional algebras, this is equivalent to the requirement of strict positivity of non-normalised density matrices that form the representation of the quantum model over the GNS Hilbert space. This excludes the possibility of consideration of pure quantum states as elements of quantum manifolds. From the geometric perspective, this can be understood as restriction of considerations to the differential manifolds without boundary, since pure states form a subset of the boundary. Thus, emergent space-times are defined only for mixed states. A point of a quantum Schwarzschild space-time that was constructed in this paper is a four-dimensional strictly positive matrix, which is a direct sum of two two-dimensional strictly positive matrices. However, these matrices are not \textit{qubits} in the usual understanding of that term, because qubits require an additional normalisation constraint, which is not satisfied by our construction. The normalisation condition reduces the dimensionality of the quantum model, so in order to construct four-dimensional space times based on qubits, one would need to use different representation of quantum models. 

%In the forthcoming paper \cite{Kostecki:Menicucci:2011}, the closely related inverse problem is solved: starting from a given quantum model and its geometry (the Bloch sphere of a qubit), the space-time that is generated by this model is identified. But w
While the choice of quantum model and its geometry determines to a large extent the corresponding space-time (it remains to chose the global foliation for a Poincar\'{e}--Wick rotation, which for some models is naturally suggested by their geometry), the inverse problem of construction of quantum model that generates a particular space-time is generally harder and it might admit many very different solutions. This is also in the case considered in this work: there might exist other quantum information models $\M(\N)$ that generate Schwarzschild space-time (either for the same or for some other choice of quantum riemannian metric). The characterisation of all quantum models $(\M(\N),\gbold,\mathbf{e})$ that generate Schwarzschild space-time manifold remains an interesting open problem.

The choice of the Bogolyubov--Kubo--Mori riemannian metric $\gbold$ on $\M(\N)$ is determined, according to (\ref{quantum.Eguchi}), by the choice of Umegaki's relative entropy functional on $\M(\N)$. Hence, as long as no other principles determining the riemannian geometries on quantum models are considered, this is the most canonical choice (from the perpective of quantum information theory). On the other hand, the particular coordinate system used in the above derivation is by no means unique. It is just a convenient tool to provide calculations required by the use of the Fronsdal embedding. Besides this particular aim, the construction of Schwarzschild space-time based on a family $Q_{fg}$ of manifolds is quite inconvenient. It would be interesting to find some other class of quantum information models generating the Schwarzschild space-time, which could be defined directly in terms of some operational (experimental) constraints. However, this would require to use of some other technique of construction of the four-dimensional manifold. The advantage of the method used in this paper is that it utilises the representation of quantum states in terms of Pauli matrices, what allows a remarkable simplification of the formula for the BKM metric. As a result, the system of differential equations generated by the flatness condition was analytically solvable.

This shows that the problem of operational meaning of quantum model that generates a particular space-time is closely related with the method used to introduce a particular riemannian metric $\gbold$ on this model. Putting it more strongly, we think that in order to justify the choice of a quantum model which generates a given space-time, one necessarily has to provide an explicit operational semantics that serves as an environment (operational context) for such choice. Without such environment, it is impossible to identify the operational meaning of the mathematical parameters of the emergent space-times (e.g., the parameter $m$ in the Schwarzschild solution (\ref{Schwarz.big})).

Both Hilbert space based kinematics of orthodox approach to quantum theory and lorentzian geometry of space-time arise as two representations of the underlying quantum information geometry of $\M(\N)$. Both amount to forgetting some part of the structure of the quantum information model. But because they have the same origin, they are mutually related from scratch. As a result, to each quantum space-time $(\M(\N),\tilde{\gbold})$ there is assigned a `classical' space-time $(\M_c,\tilde{\gbold}_c):=\mathrm{DeQuant}^{\mathrm{lor},\mathrm{to}}(\M(\N),\tilde{\gbold})$, as well as a subset $\mathcal{L}:=\ell_{1/2}(\M(\N))$ of a Hilbert space $\H\cong L_2(\N)$, such that to each element of $\M_c$ there corresponds a vector in $\mathcal{L}\subset\H$. Given a GNS representation $\pi_\omega:\N\ra\BBB(\H_\omega)$ of a finite dimensional algebra $\N$ (provided by the choice of some $\omega\in\N_*^+$), the Hilbert space $\H$ is unitarily isomorphic to $\H_{HS}=L_2(\BBB(\H_\omega),\Tr)$ and the vectors in $\mathcal{L}\subset\H$ correspond (via the inverse of $\rho\mapsto 2\rho^{1/2}$) to the non-normalised density matrices in $\BBB(\H_\omega)$. Hence, in particular, a (continuous or discrete) space-time trajectory in $\M_c$ corresponds uniquely to a family of density matrices represented as a (respectively, continuous or discrete) trajectory of vectors in $\mathcal{L}\subset\H$. %Whenever $\N$ is a type I factor (that is, within the regime of validity of quantum \textit{mechanics}), the space $\H$ is isomorphic to the space $\K\otimes\K^\banach$, where $\K$... 

This way the point of a space-time and the density matrix of quantum theory can be considered just as two representations of the single quantum state of information $\omega\in\M(\N)$. The difference between two space-time points can be identified only by specifying some difference between two quantum states of information that define these points. In this sense, the primary property of the space-time event is no longer its location in some causal poset. Causality is just a special, and emergent, case of correlativity: in general, the space-time events are distinguishable only by their correlation contents. If a particular operational semantics defining the geometric data $(\M(\N),\gbold,\mathbf{e})$ is provided, then an emergent space-time becomes a purely epistemic entity: its points and its geometry represent only some quantified knowledge, with no ontological (substantial) contents whatsoever.

Note that in this paper we discuss only the kinematic aspect of emergence of space-time from quantum theory. The choice of a particular quantum riemannian metric and global `temporal' one-form is considered as a part of a definition of quantum kinematics. The dynamical features of the relationship between quantum theory and space-time, including  trajectories representing the non-linear quantum dynamics (generated by constrained maximisation of quantum relative entropy \cite{Kostecki:2010:AIP,Kostecki:2011:Vaxjo}) and the quantum analogue of the Hilbert--Einstein variational equations, will be discussed elsewhere.

\ \\\textbf{Acknowledgements}. We would like to thank Wojtek Kami\'{n}ski and W{\l}odek Natorf for their valuable comments. This research was partially supported by MSWiN \textit{182/N} QGG/2008/0 and NCN \textit{N N202} 343640 grants. RPK thanks also Roman Kalyakin and Katya Kamlovskaya for their kind help with accommodation.

\nocite{Petz:1994:geometry}
\nocite{Petz:Toth:1993} 
\nocite{Rodriguez:1999}

{\small
\bibliographystyle{../rpkbib}
\bibliography{../rpkrefs}
}
\end{document}